\documentstyle[12pt,axodraw]{article}
\newcommand{\EPJ}{Eur. Phys. J. }

\newcommand{\JHEP}{J. High Energy Phys. }

\newcommand{\NP}{Nucl. Phys. }
\newcommand{\PR}{Phys. Rev. }
\newcommand{\PRL}{Phys. Rev. Lett. }
\newcommand{\PL}{Phys. Lett. }

\newcommand{\lnc}{\Lambda_{\rm NC}}

\begin{document}
\baselineskip=20pt

\pagenumbering{arabic}

\vspace{1.0cm}
\begin{flushright}
LU-ITP 2002/024 
\end{flushright}

\begin{center}
{\Large\sf Some phenomenological consequences of the time-ordered perturbation 
theory of QED on noncommutative spacetime}\\[10pt]
\vspace{.5 cm}

{Yi Liao, Christoph Dehne}
\vspace{1.0ex}

{\small Institut f\"ur Theoretische Physik, Universit\"at Leipzig,
\\
Augustusplatz 10/11, D-04109 Leipzig, Germany\\}

\vspace{2.0ex}

{\bf Abstract}

\end{center}

A framework was recently proposed for doing perturbation theory on 
noncommutative (NC) spacetime. It preserves the unitarity of $S$ matrix 
and differs from the naive, popular approach already at the lowest 
order in perturbation when time does not commute with space. In this 
work, we investigate its phenomenological implications at linear 
colliders, especially the TESLA at DESY, through the processes of 
$e^+e^-\to\mu^+\mu^-,H^+H^-,H^0H^0$. 
The results are indeed found to be very different from the ones 
obtained in the naive approach. The first two get corrected at tree 
level as opposed to the null result in the naive approach, while the 
third one coincides with the naive result only in the low energy 
limit. The impact of the Earth rotation is incorporated. The NC 
signals are generally significant when the NC scale is comparable to 
the collider energy. If this is not the case, the nontrivial azimuthal 
angle distribution and day-night asymmetry of events due to Lorentz 
violation and the Earth rotation will be useful in identifying signals. 
We also comment briefly on the high energy behaviour of the cross 
section that grows up linearly in the center of mass energy squared and 
argue that it does not necessarily contradict some statements, e.g., the 
Froissart-Martin bound, achieved in ordinary theory. 

\begin{flushleft}


\end{flushleft}

\newpage
\section{Introduction}

Field theory on noncommutative (NC) spacetime has stimulated a lot of 
investigations since it was found to arise naturally in the context of 
string theory $\cite{string}$. A possible way to formulate a field 
theory on NC spacetime is to implement the Weyl-Moyal correspondence 
that replaces in the action the usual product of field operators by 
their star product, 
\begin{equation}
\begin{array}{rcl}
(f_1\star f_2)(x)&=&\displaystyle 
\left[\exp\left(\frac{i}{2}
\theta^{\mu\nu}\partial^x_{\mu}\partial^y_{\nu}\right)
f_1(x)f_2(y)\right]_{y=x}. 
\end{array}
\end{equation}
Here $x,y$ are the usual commutative coordinates and 
$\theta_{\mu\nu}$ 
is a real, antisymmetric, constant matrix characterizing the 
noncommutativity of spacetime, 
$[\hat{x}_{\mu},\hat{x}_{\nu}]=i\theta_{\mu\nu}$. The $\theta$ 
parameter has the dimension of length squared and is presumably related 
to some energy scale $\lnc$ where NC physics sets in. Considering the 
connection of NC field theory to string theory and that gravity may be 
unified with gauge interactions in the string framework at a TeV scale 
$\cite{tev}$, it seems reasonable to expect that $\lnc$ should be not 
far above a TeV. This possibility opened an avenue to extensive 
phenomenological studies that could test the ideas of NC physics in low 
or high energy experiments$\cite{hewett}$-$\cite{low}$. Concerning these, 
we mention one point which will be relevant to our present work. Since 
$\theta_{\mu\nu}$ is a constant matrix instead of a Lorentz tensor, 
Lorentz invariance is explicitly broken in NC field theory. As pointed 
out in Ref. $\cite{hewett}$, this has important repercussions on data 
analysis of collider experiments done on the Earth which rotates by 
itself and 
revolves around the Sun. In Ref. $\cite{liao01}$, it has been shown how 
this seemingly troublesome problem can be used as an advantage in 
discriminating NC signals from those in the ordinary commutative theories 
like the standard model and other new physics. It has also been argued 
that for collider experiments on the Earth the main impact comes from the 
relative change of the directions of $\theta_{\mu\nu}$ to the locally fixed 
reference frame as the Earth rotates. This apparent change of directions 
has been further elaborated upon recently in Ref. $\cite{kamoshita}$. We 
shall continue to include this effect in the present work. 

All of perturbative calculations performed so far in NC field theory have 
been based on the understanding that the only difference of NC theory from 
its commutative counterpart is the appearance of NC phases in Feynman 
rules of interaction vertices $\cite{filk,others}$. 
It was found however that such 
a perturbation framework cannot preserve the unitarity of $S$ matrix when 
time does not commute with space, i.e., $\theta_{0i}\ne 0$ 
$\cite{unitarity}$. This may be understood as follows. A typical NC phase 
looks like $\exp(i/2~\theta_{\mu\nu}p^{\mu}k^{\nu})$ where $p,k$ are the 
momenta of the relevant particles. When it appears in a loop diagram, one 
of the momenta will represent the loop momentum to be integrated over. For 
$\theta_{0i}\ne 0$, the zero-th component of the loop momentum enters into 
the phase. Then we cannot arbitrarily Wick rotate it to the imaginary axis 
since there is no guarantee any more that the integrand vanishes rapidly 
enough at infinity in the complex plane. In such a case, the imaginary part 
of a forward scattering amplitude will get additional contributions from 
the NC phase. This is in sharp contradiction to the unitarity relation 
which states that the imaginary part can only be associated with physical 
thresholds which in turn are determined by internal propagators 
independently of vertices. This failure in unitarity may not be necessarily 
attributed to the noncommutativity of time and space, but may well be due 
to the improper perturbation framework instead. Indeed, as demonstrated 
for the one-loop self-energy in $\varphi^3$ theory in Ref. 
$\cite{bahns}$ (see also Ref. $\cite{rim}$), 
the approach using the Yang-Feldman equation gives a unitary result that is 
consistent with the general considerations in Ref. $\cite{doplicher}$. More 
recently, starting from some basic assumptions concerning perturbation 
theory that are commonly adopted in the literature, it has been shown 
$\cite{liao}$ (see also $\cite{bozkaya}$) 
that a careful treatment of the time-ordering procedure does 
not lead to the naive formalism as first formulated in Ref. $\cite{filk}$ 
when $\theta_{0i}\ne 0$. 
Instead, it results in a framework which is the old-fashioned, time-ordered 
perturbation theory (TOPT) 
modified properly to the NC context. Unitarity can be preserved as long as 
the interaction Lagrangian is explicitly Hermitian. More importantly, the 
new framework differs from the old one already at tree level although the 
two become identical when $\theta_{0i}=0$. The whole picture of perturbation 
theory is thus altered for the case of $\theta_{0i}\ne 0$. 

In the present work we shall start to pursue the phenomenological 
consequences of the new perturbation framework. We shall work out the 
simplest processes at linear colliders that feel NC effects, 
$e^+e^-\to\mu^+\mu^-,H^+H^-,H^0H^0$. Here $H^{\pm},H^0$ could be any scalars 
that participate in NC scalar QED interactions although they will be considered 
as Higgs bosons in our mind. We shall find in the next section that the 
results are indeed very different from those obtained in the naive approach. 
In the latter approach, there are no corrections at tree level to the first 
two processes, and the third one was computed in Ref. $\cite{liao01}$. 
However, in TOPT the first two also get corrected and the third one 
approaches the naive result only in the low energy limit. More surprisingly, 
when the process occurs in the $s$ channel through a massless intermediate 
state, the NC corrected term in the cross section can grow up linearly in 
$s$, the center of mass energy squared. We shall argue in the last section 
how this does not necessarily contradict the Froissart-Martin bound obtained 
in ordinary field theory. Our numerical analysis including the Earth rotation 
effect is detailed in section $3$, and our results are summarized in the last 
section.

\section{Calculation of the processes}

We present the analytic part of our calculation in this section. We first 
review the time-ordered perturbation theory adapted for NC field theory 
in Ref. $\cite{liao}$. This is then followed by calculation of cross 
sections in the local reference frame fixed to a particular collider. The 
Earth rotation effects are included in the last subsection.  

\subsection{Computational rules}

Ref. $\cite{liao}$ starts with some basic assumptions about perturbation 
theory on NC spacetime that are usually made in the literature. The Green 
functions are defined in terms of vacuum expectation values of the 
time-ordered products of field operators and the exponentiated interaction 
action. The usual time-ordering procedure is adopted and the free theory 
is taken to be a good starting point for perturbation theory. It has 
then been shown that, when $\theta_{0i}\ne 0$ the resulting perturbation 
framework is not the naive, seemingly covariant one as extensively used 
in the literature $\cite{filk,others}$, but the 
old-fashioned, time-ordered perturbation theory extended with new NC 
interaction vertices. The two formalisms coincide when time commutes with 
space, but are not equivalent in the opposite case especially concerning 
the fate of unitarity. In the language of 
TOPT, a physical process is virtualized as a series of transitions 
between physical intermediate states that are sequential in time. The 
unitarity of $S$ matrix is thus apparent if the interaction Lagrangian 
is explicitly Hermitian. Actually, as shown there, unitarity holds valid 
for generally off-shell Green functions as well. In the following we 
shall list the computational rules for NC vertices to be used here which 
are part of the prescriptions spelled out in Ref. $\cite{liao}$. We refer 
the interested reader to that reference for a detailed exposition.

For our purpose, the Lagrangian for spinor and scalar QED on NC spacetime 
is given by, 
\begin{equation}
\begin{array}{rcl}
{\cal L}&=&\displaystyle -\frac{1}{4}F^{\mu\nu}\star F_{\mu\nu}
+\bar{\psi}\star(\gamma^{\mu}iD_{\mu}-m)\psi\\
&&\displaystyle +\frac{1}{2}(D_{\mu}\varphi_0)\star(D^{\mu}\varphi_0)
-\frac{1}{2}m_0^2\varphi_0\star\varphi_0\\
&&\displaystyle +(D_{\mu}\varphi)^{\dagger}\star(D^{\mu}\varphi)
-m_{\pm}^2\varphi^{\dagger}\star\varphi.
\end{array}
\end{equation}
Here $\psi$ is the charged spinor field ($e^-$ or $\mu^-$) with mass $m$, 
$\varphi$ and $\varphi_0$ are the charged ($H^-$) and neutral ($H^0$) 
scalar fields with mass $m_{\pm}$ and $m_0$ respectively. $A$ is the 
electromagnetic field with coupling $e$. The covariant derivatives and 
field tensor are defined by the generalized gauge invariance, 
\begin{equation}
\begin{array}{rcl}
F_{\mu\nu}&=&\displaystyle
\partial_{\mu}A_{\nu}-\partial_{\nu}A_{\mu}
+ie[A_{\mu},A_{\nu}]_{\star},\\
D_{\mu}\varphi_0&=&\displaystyle\partial_{\mu}\varphi_0
+ie[A_{\mu},\varphi_0]_{\star},\\
D_{\mu}\varphi&=&\displaystyle\partial_{\mu}\varphi
+ieA_{\mu}\star\varphi,\\
D_{\mu}\psi&=&\displaystyle\partial_{\mu}\psi
+ieA_{\mu}\star\psi,
\end{array}
\end{equation}
where $[A,B]_{\star}=A\star B-B\star A$ is the Moyal bracket. The 
processes to be considered here also occur through weak interactions 
in the standard model, so in principle we should include the NC 
modifications for this part. Since the attempts to generalize the 
standard model to NC spacetime have not been successful so far 
$\cite{nogo}$, 
we shall not try to guess what the modified weak interactions would 
look like. In this sense, our calculations should be considered as 
NC corrections to QED results. 

For the processes here, we only need the rules for the vertices 
$A\bar{\psi}\psi, A\varphi^{\dagger}\varphi$ and 
$A\varphi_0\varphi_0$, which are found to be, 
\begin{equation}
\begin{array}{rcl}
\bar{\psi}(p_2,\lambda_2)\psi(p_1,\lambda_1)A^{\mu}(p_3,\lambda_3)
&=&\displaystyle 
+e\gamma^{\mu}P_{231},\\
\varphi^+(p_2,\lambda_2)\varphi(p_1,\lambda_1)A^{\mu}(p_3,\lambda_3)
&=&\displaystyle 
+e(p_{1\lambda_1}-p_{2\lambda_2})^{\mu}P_{231},\\
\varphi^0(p_2,\lambda_2)\varphi^0(p_1,\lambda_1)A^{\mu}(p_3,\lambda_3)
&=&\displaystyle 
-e/2[p_{1\lambda_1}^{\mu}(P_{132}-P_{231}+P_{321}-P_{123})\\
&&\displaystyle 
~~~~+p_{2\lambda_2}^{\mu}(P_{231}-P_{132}+P_{312}-P_{213})], 
\end{array}
\end{equation}
where the NC phase 
$P_{jk\ell}=\exp[-i(p_{j\lambda_j},p_{k\lambda_k},
p_{\ell\lambda_{\ell}})]$ with $(a,b,c)=a\wedge b+b\wedge c+a\wedge c$ 
and $a\wedge b=1/2~\theta_{\mu\nu}a^{\mu}b^{\nu}$. All momenta 
are meant to be incoming. The parameter $\lambda$ denotes the direction 
of time flow along the momentum direction, which is 
$+(-)$ if the vertex is the later (earlier) end of the line. 
$p_{\lambda}^{\mu}=(\lambda E_{\bf{p}},\bf{p})$ with 
$E_{\bf{p}}=\sqrt{\bf{p}^2+\mu^2}$ 
denotes the on-shell four-momentum with positive or negative energy of a 
particle with mass $\mu$ and three-momentum $\bf{p}$. It should be noted 
that only on-shell momenta are involved in the vertices. This is indeed 
in the spirit of TOPT which treats all intermediate states as physical 
ones. The point here, as emphasized in Ref. $\cite{liao}$ is that, the 
seemingly covariant formalism which treats intermediate states as off-shell 
cannot be recovered as in ordinary theory when $\theta_{0i}\ne 0$. This 
will also be verified in our following calculations.  

\subsection{Cross sections in the locally fixed reference frame}

Let us first compute the process of $e^+e^-\to\mu^+\mu^-$. Its 
time-ordered Feynman diagrams are shown in Fig. $1$, where the wavy and 
arrowed solid lines stand for the photon and $e^-$ (or $\mu^-$) fields 
respectively. The other arrows indicate momenta and the $\lambda$ 
parameters of the time flow. For on-shell scattering we have 
$\lambda_j=+$ for all external particles ($j=1,2,3,4$) while 
$\lambda=+,-$ corresponds to the two 
possible time-flows of the intermediate photon shown in Fig. $1(a)$ and 
$1(b)$. 

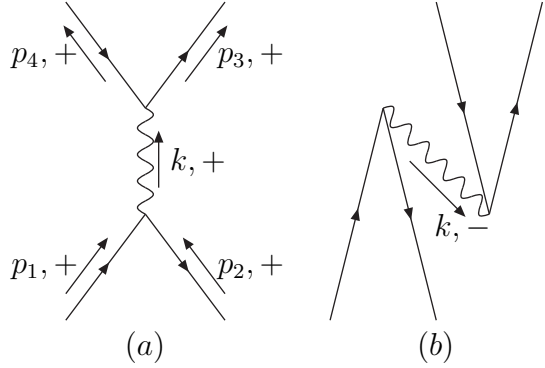
\begin{figure}
\begin{center}
\begin{picture}(250,150)(0,0)
\SetOffset(50,30)
\ArrowLine(0,0)(30,40)\ArrowLine(30,40)(60,0)
\LongArrow(0,10)(15,30)\LongArrow(60,10)(45,30)
\ArrowLine(0,120)(30,80)\ArrowLine(30,80)(60,120)
\LongArrow(15,90)(0,110)\LongArrow(45,90)(60,110)
\Photon(30,40)(30,80){3}{4}
\LongArrow(35,50)(35,70)
\Text(-8,20)[]{$p_1,+$}\Text(70,20)[]{$p_2,+$}
\Text(-8,100)[]{$p_4,+$}\Text(70,100)[]{$p_3,+$}
\Text(50,60)[]{$k,+$}
\Text(30,-10)[]{$(a)$}

\SetOffset(150,30)
\ArrowLine(0,0)(20,80)\ArrowLine(20,80)(40,0)
\ArrowLine(40,120)(60,40)\ArrowLine(60,40)(80,120)
\Photon(20,80)(60,40){3}{6}
\LongArrow(30,60)(50,40)\Text(50,35)[]{$k,-$}
\Text(40,-10)[]{$(b)$}

\SetOffset(0,0)
\end{picture}
\end{center}
\caption{Time-ordered diagrams for $e^+e^-\to\mu^+\mu^-$. Time 
flows upwards.}
\end{figure}

Using the prescriptions in Ref. $\cite{liao}$ and the above rules 
for vertices, the sum of the two diagrams is, 
\begin{equation}
\begin{array}{l}
\displaystyle
\bar{v}_2e\gamma_{\mu}u_1~\bar{u}_3e\gamma^{\mu}v_4
~(-2\pi)\delta(E_1+E_2-E_3-E_4)\sum_{\lambda}
\int\frac{d^3\bf{k}}{(2\pi)^3 2\omega_{\bf{k}}}\\
\displaystyle \times 
(2\pi)^3\delta^3({\bf p}_1+{\bf p}_2-{\bf k})^3
(2\pi)^3\delta^3({\bf k}-{\bf p}_3-{\bf p}_4)^3
(-1)[\lambda(E_1+E_2)-\omega_{\bf k}+i\epsilon]^{-1}\\
\displaystyle \times 
\exp[-i(p_{2\lambda_2},-k_{\lambda},p_{1\lambda_1})]
\exp[-i(-p_{3\lambda_3},k_{\lambda},-p_{4\lambda_4})], 
\end{array}
\end{equation}
where $E_j$'s are the external particles energies and 
$\omega_{\bf k}=|{\bf k}|$ is the energy of the intermediate photon. 
Note that there is an extra minus sign for the energy deficit factor 
for the photon intermediate state compared to the scalar one. Finishing 
the phase space integral and dropping the global four-momentum 
conservation factor, we obtain the transition amplitude as 
\begin{equation}
\begin{array}{rcl}
{\cal A}&=&\displaystyle 
\bar{v}_2e\gamma_{\mu}u_1~\bar{u}_3e\gamma^{\mu}v_4
(2\omega_{\bf{k}})^{-1}\sum_{\lambda}
[\lambda(E_1+E_2)-\omega_{\bf k}+i\epsilon]^{-1}\\
&\times&\displaystyle 
\exp[-i(p_{2+},-k_{\lambda},p_{1+})]
\exp[-i(p_{3+},-k_{\lambda},p_{4+})],
\end{array}
\label{eq_mu}
\end{equation}
with $k=p_1+p_2=p_3+p_4$. The plus sign 
subscripts to $p_j$'s may be dropped since they are already on-shell 
four-momenta of positive energy. We keep them to emphasize the point 
that, generally, 
$p_{1\lambda_1}+p_{2\lambda_2}\ne k_{\lambda}\ne
p_{3\lambda_3}+p_{4\lambda_4}$. 

When $\theta_{0i}=0$, all $\lambda$ parameters are automatically removed 
from the NC phases. Then the $\lambda$ dependence resides exclusively in 
the energy deficit factor. The NC phases may be simplified using the 
four-momentum conservation, e.g., 
$\exp[-i(p_2,-k,p_1)]=\exp[-ip_1\wedge p_2]$. The sum over $\lambda$ is 
finished using, 
\begin{equation}
\frac{1}{2\omega_{\bf{k}}}
\sum_{\lambda}\frac{1}{\lambda(E_1+E_2)-\omega_{\bf k}+i\epsilon}
=\frac{1}{k^2+i\epsilon}. 
\end{equation}
We thus reproduce the result that would have been obtained in the naive 
approach, 
\begin{equation}
\begin{array}{rcl}
{\cal A}&=&\displaystyle 
\bar{v}_2e\gamma_{\mu}u_1~\bar{u}_3e\gamma^{\mu}v_4
\frac{e^{-ip_1\wedge p_2}e^{-ip_4\wedge p_3}}{k^2+i\epsilon}, 
{\rm ~for~}\theta_{0i}=0, 
\end{array}
\end{equation}
which deviates from the QED result by a global phase only and thus gives 
the same cross section.  

For $\theta_{0i}\ne 0$ however, the $\lambda$ dependence remains in NC 
phases. Denoting $n^{\mu}=(0,\hat{\bf k})$ so that 
$k^{\mu}_{\lambda}=\lambda\omega_{\bf k}(1,\vec{0})+\omega_{\bf k}n^{\mu}$, 
we have for any momentum $q$, 
$k_{\lambda}\wedge q=\omega_{\bf k}(1/2~\lambda\theta_{0i}q^i+n\wedge q)$.
The NC phases in eq. $(\ref{eq_mu})$ become, 
\begin{equation}
\exp[-i(p_2\wedge p_1+p_3\wedge p_4+\omega_{\bf k}n\wedge q)]
\exp[-i\lambda\omega_{\bf k}\theta_{0j}q^j/2], 
\end{equation}
with $q=p_2+p_3-p_1-p_4$. The sum over $\lambda$ in eq. $(\ref{eq_mu})$ can 
be finished using 
\begin{equation}
\displaystyle\frac{1}{2\omega_{\bf{k}}}\sum_{\lambda}
\frac{\exp[-i\lambda a]}{\lambda(E_1+E_2)-\omega_{\bf k}+i\epsilon}
=\displaystyle\frac{1}{k^2+i\epsilon}
\left[\cos a-i(E_1+E_2)\frac{\sin a}{\omega_{\bf k}}\right], 
\label{eq_sum}
\end{equation}
with $a=\omega_{\bf k}\theta_{0j}q^j/2$. 
Note that in contrast to the case of $\theta_{0j}=0$, the NC phases for the 
two time-flows do not factorize any more; instead, they interfere to produce 
a new imaginary part. Then, 
\begin{equation}
\begin{array}{rcl}
{\cal A}&=&\displaystyle 
\exp[-i(p_2\wedge p_1+p_3\wedge p_4+\omega_{\bf k}n\wedge q)]
~\bar{v}_2e\gamma_{\mu}u_1~\bar{u}_3e\gamma^{\mu}v_4\\
&\times&\displaystyle \frac{1}{k^2+i\epsilon}
\left[\cos a-i(E_1+E_2)\frac{\sin a}{\omega_{\bf k}}\right]. 
\end{array}
\end{equation}

Now we specialize to the center of mass frame (c.m.) fixed locally to a 
particular collider. Then, ${\bf k},\omega_{\bf k}\to 0$ so that 
\begin{equation}
\begin{array}{rcl}
{\cal A}&=&\displaystyle 
\exp[-i(p_2\wedge p_1+p_3\wedge p_4)]
~\bar{v}_2e\gamma_{\mu}u_1~\bar{u}_3e\gamma^{\mu}v_4\\
&\times&\displaystyle \frac{1}{k^2+i\epsilon}
\left[1-i\frac{1}{2}(E_1+E_2)\theta_{0j}q^j\right]\\ 
&=&\displaystyle 
\exp[-i(p_2\wedge p_1+p_3\wedge p_4)]
~\bar{v}_2e\gamma_{\mu}u_1~\bar{u}_3e\gamma^{\mu}v_4\\
&\times&\displaystyle s^{-1}
\left[1+i\theta_{0j}(p_1-p_3)^j\sqrt{s}\right], {\rm ~in~c.m.~}, 
\end{array}
\end{equation}
where $s$ is the c.m. energy squared. The unpolarized differential 
cross section is found to be, 
\begin{equation}
\begin{array}{rcl}\displaystyle 
\frac{d\sigma^{\mu}}{d\Omega}&=&\displaystyle 
\frac{\alpha^2}{4s}(1+c^2_{\theta})f^{\mu}, 
\end{array}
\end{equation}
where $c_{\theta}=\cos\theta,s_{\theta}=\sin\theta$ etc with 
$\theta,\varphi$ being the polar and azimuthal angles of $\mu^-$ in the local 
reference frame. (This $\theta$ should not be confused with the NC 
parameter.) $f^{\mu}$ is the NC correction factor, 
\begin{equation}
\begin{array}{rcl}
f^{\mu}&=&\displaystyle 1+\frac{1}{4}
\left(\frac{\sqrt{s}}{\lnc}\right)^4(w_i-w_f)^2, 
\end{array}
\end{equation}
where we have defined the NC parameter vector $\theta_E^j=\theta^{0j}_E$, 
with $\lnc=|\vec{\theta}_E|^{-1/2}$ being the associated NC energy scale, 
and 
$w_i=\hat{\theta}_E\cdot\hat{\bf p}_1, 
w_f=\hat{\theta}_E\cdot\hat{\bf p}_3$ using the relevant unit vectors. 
We have also ignored the masses of the electron and muon which is 
appropriate for high energy collisions. Note that the NC correction 
term in $f^{\mu}$ grows up as $s^2$. Since we did not make any low energy 
approximations, it holds true at high energies as long as the NC QED 
persists to be a valid description of QED interactions on NC spacetime. 
This phenomenon arises due to the balance of the two competing factors. 
While the exchange of a soft, massless and on-shell photon in the center 
of mass tends to blow up the amplitude, the opposite but small NC phases 
from the two time-flows tend to annihilate the contributions. We shall 
discuss in the last section how this high energy behaviour does not 
necessarily contradict the Froissart-Martin bound derived in ordinary 
theory. Here we just comment that it implies a lower bound on the 
differential cross section, 
\begin{equation}
\begin{array}{rcl}\displaystyle 
\frac{d\sigma^{\mu}}{d\Omega}&\ge&\displaystyle 
\frac{\alpha^2}{4\lnc^2}(1+c^2_{\theta})|w_i-w_f|. 
\end{array}
\end{equation}

The $H^+(p_4)H^-(p_3)$ production is similarly computed. Using the same 
notation as above, we find, 
\begin{equation}
\begin{array}{rcl}
{\cal A}&=&\displaystyle 
\exp[-i(p_2\wedge p_1+p_3\wedge p_4+\omega_{\bf k}n\wedge q)]
~e^2\bar{v}_2(\rlap/p_3-\rlap/p_4)u_1\\
&\times&\displaystyle \frac{1}{k^2+i\epsilon}
\left[\cos a-i(E_1+E_2)\frac{\sin a}{\omega_{\bf k}}\right], 
\end{array}
\end{equation}
which again reduces to the standard QED result up to a global phase as 
in the naive approach when $\theta_{0j}=0$. For $\theta_{0j}\ne 0$ 
however, the second term also contributes. In the c.m., we have, 
\begin{equation}
\begin{array}{rcl}
{\cal A}&=&\displaystyle 
\exp[-i(p_2\wedge p_1+p_3\wedge p_4)]
~e^2\bar{v}_2 2\rlap/p_3u_1\\
&\times&\displaystyle s^{-1}
\left[1+i\theta_{0j}(p_1-p_3)^j\sqrt{s}\right],\\
\displaystyle 
\frac{d\sigma^{H^{\pm}}}{d\Omega}&=&\displaystyle 
\frac{\alpha^2}{8s}\beta^3_{\pm}s^2_{\theta}f^{H^{\pm}}, 
\end{array}
\end{equation}
where $\beta_{\pm}=\sqrt{1-4m_{\pm}^2/s}$ is the final particle velocity and, 
\begin{equation}
\begin{array}{rcl}
f^{H^{\pm}}&=&\displaystyle 1+\frac{1}{4}
\left(\frac{\sqrt{s}}{\lnc}\right)^4(w_i-\beta w_f)^2. 
\end{array}
\end{equation}

The $H^0(p_3)H^0(p_4)$ process also occurs at tree level through an 
$s$-channel exchange of photon in NC QED. Its computation is slightly 
more complicated but straightforward. Finishing the trivial intermediate 
state phase space integral, we obtain, 
\begin{equation}
\begin{array}{rcl}
{\cal A}&=&\displaystyle 
\frac{1}{2}e^2\bar{v}_2\rlap/p_3u_1
e^{-ip_2\wedge p_1}\sum_{\lambda}\frac{g(k_{\lambda})}
{2\omega_{\bf k}[\lambda(E_1+E_2)-\omega_{\bf k}+i\epsilon]}, 
\end{array}
\end{equation}
where 
\begin{equation}
\begin{array}{rcl}
g(k_{\lambda})&=&\displaystyle 
e^{-ip_3\wedge p_4}[2e^{i2\omega_{\bf k}n\wedge(p_1-p_3)}
e^{i\lambda\omega_{\bf k}\theta_{0j}(p_1-p_3)^j}\\
&&\displaystyle 
-e^{-i2\omega_{\bf k}n\wedge p_2}
e^{-i\lambda\omega_{\bf k}\theta_{0j}p_2^j}
-e^{i2\omega_{\bf k}n\wedge p_1}
e^{i\lambda\omega_{\bf k}\theta_{0j}p_1^j}]\\
&&\displaystyle 
-(p_3\leftrightarrow p_4).
\end{array}
\end{equation}
Note that ${\cal A}$ is symmetric in $p_3$ and $p_4$ since the spinor 
part is antisymmetric as is $g(k_{\lambda})$. 
The summation over $\lambda$ can be finished as in eq. $(\ref{eq_sum})$, 
which is essentially, 
\begin{equation}
\displaystyle 
\sum_{\lambda}\frac{g(k_{\lambda})}
{2\omega_{\bf k}[\lambda(E_1+E_2)-\omega_{\bf k}+i\epsilon]} 
=\frac{1}{s}\left[\frac{1}{2}\left(g(k_+)+g(k_-)\right)+(E_1+E_2)
\frac{g(k_+)-g(k_-)}{2\omega_{\bf k}}\right].
\end{equation}
For $\theta_{0j}=0$, the above sum reduces to 
$s^{-1}i8\sin(p_3\wedge p_4)$ so that 
\begin{equation}
\begin{array}{rcl}
{\cal A}&=&\displaystyle 
i4e^2e^{ip_2\wedge p_1}\sin(p_3\wedge p_4)s^{-1}\bar{v}_2\rlap/p_3u_1, 
{\rm ~for~}\theta_{0j}=0, 
\end{array}
\label{eq_naive}
\end{equation}
which is the result obtained in the naive approach $\cite{liao01}$. The 
result for $\theta_{0j}\ne 0$ is lengthy, so we specialize to the c.m., 
\begin{equation}
\begin{array}{rcl}
{\cal A}&=&\displaystyle 
-i2e^2e^{-ip_2\wedge p_1}\bar{v}_2\rlap/p_3u_1s^{-1}(E_1+E_2)
\theta_{0j}p_3^j\cos[(E_1+E_2)\theta_{0j}p_3^j/2], 
\end{array}
\end{equation}
which coincides with the naive one eq. $(\ref{eq_naive})$ only to the 
leading order in the low energy limit. The differential cross section is 
\begin{equation}
\begin{array}{rcl}
\displaystyle\frac{d\sigma^{H^0}}{d\Omega}&=&\displaystyle
\frac{\alpha^2}{64}\frac{s}{\lnc^4}\beta^5_0s_{\theta}^2 f^{H^0},\\
\displaystyle f^{H^0}&=&\displaystyle  
\left[w_f\cos\left(\frac{s}{8\lnc^2}\beta_0 w_f\right)\right]^2,
\end{array}
\end{equation}
where now $\beta_0=\sqrt{1-4m^2_0/s}$. 

\subsection{Earth rotation effects}

The above results for cross sections would be the final ones to be used 
for data analysis if the collider were fixed relative to the reference 
frame in which $\theta_{\mu\nu}$'s are assigned values. This reference 
frame may be presumably defined by the microwave 
background radiation, in which a collider fixed on the Earth is constantly 
moving due to the self-rotation of the Earth and its revolution around the 
Sun. Since a collider measurement takes much longer than a day to 
accumulate data, the motion of the Earth has to be taken into account in 
data analysis. As argued in Ref. $\cite{liao01}$, the dominant effect for 
a collider experiment comes from the relative change of the preferred 
directions defined by $\theta_{\mu\nu}$ to our local terrestrial frame 
fixed by the geographic configuration of the considered collider. We shall 
include this effect in this subsection. 

We first define the celestial and terrestrial reference frames. All 
frames are assumed to be right-handed and all azimuthal-like angles are 
measured counter-clockwise. The celestial frame may be fixed to good 
precision by specifying its $3$-direction along the Earth's rotation axis 
and its $1$-direction pointing to the vernal equinox $\cite{kamoshita}$. 
The NC unit vector $\hat{\theta}_E$ is 
then measured by the polar and azimuthal angles $\rho\in[0,\pi]$ and 
$\xi\in[0,2\pi)$ which are NC physical parameters in addition to $\lnc$. 
For the local terrestrial frame, we define the $e^-$ beam as the 
$z$-direction and the (outgoing) normal to the sphere of the Earth at the 
interaction point as the $y$-direction. The polar and azimuthal angles of 
a particle momentum in this frame are denoted as $\theta\in[0,\pi]$ and 
$\varphi\in[0,2\pi)$ as in the last subsection. To consider the Earth 
rotation effect, we also have to know the geographic configuration of the 
collider which may be specified by three parameters. The site of the 
laboratory is determined by its latitude $\sigma\in[-\pi/2,+\pi/2]$ with 
positive (negative) values denoting the northern (southern) hemisphere, 
and its longitude $\omega\in[0,2\pi)$ measured, e.g., with respect to the 
vernal equinox. We assume that the direction of the $e^-$ beam deviates 
from the local meridian direction by an angle $\delta\in[0,2\pi)$. The 
angles $\sigma$ and $\delta$ are fixed parameters for each collider, while 
$\omega$ changes periodically as the Earth rotates. 

The remaining task now is to express in the local frame the fixed unit 
vector $\hat{\theta}_E$, which is accomplished using the 
standard vector analysis $\cite{kamoshita}$, 
\begin{equation}
\begin{array}{rcl}
\hat{\theta}_E^x&=&\displaystyle 
s_{\rho}(c_{\omega-\xi}s_{\sigma}s_{\delta}+s_{\omega-\xi}c_{\delta})
-c_{\rho}c_{\sigma}s_{\delta},\\
\hat{\theta}_E^y&=&\displaystyle 
s_{\rho}c_{\omega-\xi}c_{\sigma}
+c_{\rho}s_{\sigma},\\
\hat{\theta}_E^z&=&\displaystyle 
s_{\rho}(-c_{\omega-\xi}s_{\sigma}c_{\delta}+s_{\omega-\xi}s_{\delta})
+c_{\rho}c_{\sigma}c_{\delta}, 
\end{array}
\end{equation}
where $c_a=\cos a,s_a=\sin a$ for all angles. In the following we shall 
denote $\omega-\xi$ as $\omega$ for simplicity. This means effectively 
that the angle $\omega$ is measured with respect to the plane spanned by 
$\hat{\theta}_E$ and the Earth rotation axis. Then, the quantities needed 
for cross sections and defined in the last subsection are, 
\begin{equation}
\begin{array}{rcl}
w_i&=&\displaystyle \hat{\theta}_E^z,\\
w_f&=&\displaystyle \hat{\theta}_E^x s_{\theta}c_{\varphi}
+\hat{\theta}_E^y s_{\theta}s_{\varphi}+\hat{\theta}_E^z c_{\theta}. 
\end{array}
\end{equation}
The Earth rotation effect enters through the apparent change of 
$\hat{\theta}_E$ and thus cross sections. 

\section{Numerical results}

Before we present our numerical results, let us summarize the sets of 
angles introduced so far: the local angles $\theta$ and $\varphi$, the NC 
angular parameters $\rho$ and $\xi$, the configuration angles $\sigma$ and 
$\delta$ of the collider, and the Earth rotation angle $\omega$. 
Since $\xi$ appears in the combination of $\omega-\xi$, measuring $\xi$ 
amounts to setting an absolute origin for $\omega$ which may be chosen as 
the vernal equinox as mentioned above. In the following, our $\omega$ 
will be measured with respect to $\xi$ so that we shall concentrate on 
the single NC angle $\rho$. Our numerical results will be presented for 
the TESLA at DESY whose configuration angles are, 
$\sigma=53^{\circ}34^{\prime}50^{\prime\prime}$, 
$\delta=27.5^{\circ}$ $\cite{zerwas}$. 

Upon considering the Earth rotation, the differential cross sections 
computed in the last section depend on the angles $\theta,\varphi$ and 
$\omega$ as well as others. We thus may have two types of distributions, 
one in the local angles, the other in the Earth rotation angle. The 
differential cross sections can be cast collectively in the form, 
\begin{equation}
\displaystyle\frac{4\pi}{\sigma_0^A}\frac{d\sigma^A}{d\Omega}=
F^A(\theta,\varphi;\omega),~A=\mu,H^{\pm},H^0, 
\end{equation}
where $\sigma_0^{\mu,H^{\pm}}$ are the standard QED total cross sections  
while $\sigma_0^{H^0}$ is a convenient normalization constant for the 
$H^0H^0$ production, 
\begin{equation}
\displaystyle\sigma_0^{\mu}=\frac{4\pi}{3}\frac{\alpha^2}{s},~
\displaystyle\sigma_0^{H^{\pm}}=
\frac{\pi}{3}\frac{\alpha^2}{s}\beta^3_{\pm},~
\displaystyle\sigma_0^{H^0}=
\frac{\pi\alpha^2}{60}\frac{s}{\lnc^4}\beta^5_0. 
\end{equation}
Then, in terms of $f$ functions introduced in the last section, we have, 
\begin{equation}
\begin{array}{rcl}
\displaystyle 
F^{\mu}(\theta,\varphi;\omega)&=&\displaystyle 
\frac{3}{4}(1+c_{\theta}^2)f^{\mu},\\ 
\displaystyle 
F^{H^{\pm}}(\theta,\varphi;\omega)&=&\displaystyle 
\frac{3}{2}s_{\theta}^2f^{H^{\pm}},\\ 
\displaystyle 
F^{H^0}(\theta,\varphi;\omega)&=&\displaystyle 
\frac{15}{4}s_{\theta}^2f^{H^0}. 
\end{array}
\end{equation}

Let us first present the results of total cross sections or their ratios to 
the QED counterparts, which are averaged over the Earth rotation, 
\begin{equation}
\displaystyle 
\bar{\sigma}^A=\sigma_0^A\bar{R}^A,~ 
\bar{R}^A=\int\frac{d\omega}{2\pi}\int\frac{d\Omega}{4\pi}
F^A(\theta,\varphi;\omega). 
\end{equation}
For the $\mu^+\mu^-$ and $H^+H^-$ processes the integrals can be finished 
but the results are too tedious to be recorded here. In Fig. $2$ we show 
$\bar{R}^{\mu}$ as a function of $\sqrt{s}$ for different values of $\lnc$ 
and $\rho$. The NC corrections are always positive and depend strongly on 
$\lnc$ as it is clear from the form factor $f^{\mu}$. The $\rho$ dependence 
is relatively much milder. Fig. $3$ displays 
$\bar{R}^{H^{\pm}}$ as a function of mass for the chosen parameters. From 
these two plots it is clear that the viability to detect NC deviations from 
QED in total cross sections will be decisively determined by the relative 
order of magnitudes of $\lnc$ and $\sqrt{s}$. In this respect, the $H^0$ 
pair production shown in Fig. $4$ is more advantageous $\cite{liao01}$ 
since it occurs in the standard model at one loop. For comparable $\lnc$ 
and $\sqrt{s}$, e.g., $\lnc=\sqrt{s}=1$ TeV, the NC QED induced cross 
section well exceeds the one in the standard model which is about 
$0.1\sim 0.2$ fb for an intermediate mass $H^0$ $\cite{eehh}$. 

The normalized and averaged distributions in the local angles shown in 
Figs. $5$ and $6$ are defined as, 
\begin{equation}
\begin{array}{rcl}
F^A(\theta)&=&\displaystyle
\int\frac{d\omega}{2\pi}\int\frac{d\varphi}{2\pi}
F^A(\theta,\varphi;\omega),\\ 
F^A(\varphi)&=&\displaystyle
\int\frac{d\omega}{2\pi}\frac{1}{2}\int d\cos\theta~ 
F^A(\theta,\varphi;\omega). 
\end{array}
\end{equation}
Again, for comparable $\lnc$ and $\sqrt{s}$, there are sizable NC 
deviations in $\theta$ dependence from the QED results for the 
$\mu^+\mu^-$ and $H^+H^-$ production or from the standard model result 
for the $H^0H^0$ production which follows roughly the $\sim\sin^2\theta$ 
law $\cite{eehh}$. More interesting are the $\varphi$ distributions shown 
in Fig. $6$, which occurs due to the violation of Lorentz invariance. The 
distributions are also more sensitive to the $\rho$ parameter compared to 
other quantities shown above. 

The above results are obtained by time-averaging and thus correspond to 
the standard data analysis for the collider measurement. But the novel 
feature of Lorentz violation can be better displayed using the 
day-night asymmetry $\cite{liao01}$, which describes the impact of the 
Earth rotation and is defined as a function of $\omega~(\pi)$ or 
$t=12\omega/\pi$ (hour), 
\begin{equation}
\begin{array}{rcl}
A_{\rm DN}^A(\omega_a,\omega_b)&=&\displaystyle\frac
{\displaystyle\left[\int_{\omega_a}^{\omega_b}d\omega-
\int_{\omega_a+\pi}^{\omega_b+\pi}d\omega\right]\sigma^A(\omega)}
{\displaystyle\left[\int_{\omega_a}^{\omega_b}d\omega+
\int_{\omega_a+\pi}^{\omega_b+\pi}d\omega\right]\sigma^A(\omega)}\\
&=&\displaystyle\frac
{\displaystyle\left[\int_{\omega_a}^{\omega_b}d\omega-
\int_{\omega_a+\pi}^{\omega_b+\pi}d\omega\right]
\int\frac{d\Omega}{4\pi}F^A(\theta,\varphi;\omega)}
{\displaystyle\left[\int_{\omega_a}^{\omega_b}d\omega+
\int_{\omega_a+\pi}^{\omega_b+\pi}d\omega\right]
\int\frac{d\Omega}{4\pi}F^A(\theta,\varphi;\omega)}. 
\end{array}
\end{equation}
The asymmetry is shown in Fig. $7$ as histograms binned per half an 
hour for the $H^+H^-$ and $H^0H^0$ production for the same set of 
parameters as in Fig. $5$. The overall asymmetries accumulated for 
$24$ hours are respectively, 
$A_{\rm DN}^{H^{\pm}}(0,\pi)=+2.74\times 10^{-2}$ and 
$A_{\rm DN}^{H^0}(0,\pi)=-9.18\times 10^{-2}$.
This asymmetry and the azimuthal angle distribution are the most 
sensitive probe to NC signals and its angular parameter $\rho$.

\section{Conclusion and discussion}

The naive approach of perturbative NC field theory $\cite{filk,others}$ 
was shown to lead to the violation of unitarity when time does not 
commute with space $\cite{unitarity}$. This failure has been attributed 
to the improper implementation of perturbation theory $\cite{bahns}$. 
Recently, it has been demonstrated that the usual assumptions about 
perturbation theory when handled properly actually result in an 
old-fashioned, time-ordered perturbation theory modified appropriately to 
include 
effects of NC spacetime $\cite{liao}$. It turns out that this framework 
does not recover the naive one when time does not commute with space, and 
that it preserves unitarity as long as the Lagrangian is explicitly 
Hermitian. The picture for perturbation theory is thus altered; in 
particular, the difference appears already at the first order in 
perturbation. 

In the present work we investigated the phenomenological implications of 
the above framework at a high energy linear collider. We worked out the 
processes of $e^+e^-\to \mu^+\mu^-,H^+H^-,H^0H^0$ and included the effects 
of the Earth rotation. The results are indeed found to be very different 
from those obtained in the naive approach. The first two processes get 
corrected already at tree level as opposed to the naive result that 
amplitudes are only modified by a global phase and thus cross sections 
remain untouched. For the third process, the results in the two 
approaches are also different although they coincide in the low energy 
limit. The numerical significance of NC effects depends on the geographic 
configuration of the collider as well as the basic NC parameters. For 
definiteness, we presented our numerical results for the TESLA project. 
Generally speaking, the new effects are significant when the NC energy 
scale is comparable to the collider energy. When this is not the case, 
the relatively rare signals can be compensated for by their unique 
charateristics due to Lorentz violation, as shown in the azimuthal 
distribution and day-night asymmetry of events. 

Finally, we comment briefly on the surprising result on the high energy 
behaviour of cross section. We found in section $2$ that the NC correction 
term in the total cross section grows up linearly in $s$ when a soft, 
massless, on-shell photon is exchanged in the $s$-channel. Actually the 
phenomenon occurs when the mass of the exchanged particle is much less 
than the NC energy scale and $s$. We argue below that this does not 
necessarily contradict the statements in ordinary theory. For example, 
the Froissart-Martin bound that the cross section of two body reactions 
cannot grow faster than $\ln^2s$ as $s\to\infty$, was originally obtained 
$\cite{froissart}$ on the assumption of the Mandelstam's representation 
$\cite{mandelstam}$, i.e., the double dispersion relation. 
Later this bound was derived $\cite{martin}$ from the usual axioms of 
quantum field theory based on the analyticity properties of scattering 
amplitudes, in particular causality and relativistic invariance. 
In NC field theory, these last properties are 
already changed even with space-space noncommutativity alone: only a 
weaker microcausality is possible in the sense of perturbation theory at 
least and a part of Lorentz invariance survives $\cite{dispersion}$. The 
analyticity properties are altered so significantly that no dispersion 
relations have been shown to be possible for the simplest case of 
scattering $\cite{dispersion}$, let alone the double dispersion 
relations of Mandelstam. As the time-space NC is generally believed to 
be more delicate than the space-space NC, it is far from obvious that 
the Froissart-Martin bound would still apply to the time-space NC case. 
Another statement that is often made in ordinary theory is that unitarity 
sets a bound on the total cross section in the high energy limit. This is 
not a precise statement in fact. Indeed, unitarity of the $S$ matrix sets 
a bound on each of the partial wave cross sections, but it is insufficient 
to do so on the total cross section which is an infinite sum of the 
partial wave cross sections $\cite{newton}$. Only when a process is known 
to occur for a finite number of partial waves in the high energy 
limit, a bound becomes possible on the total cross section. As there are 
preferred directions in NC field theory, this partial wave analysis would 
be very different but seems to deserve further study.

\vspace{0.5cm}
\noindent
{\bf Acknowledgements} 
We thank K. Sibold for helpful discussions and carefully reading the 
manuscript. Y.L. thanks P. M. Zerwas for communications about the TESLA 
project and M. Chaichian for clarifying comments and discussions about the 
Froissart-Martin bound and drawing Ref. $\cite{martin}$ to his attention.

\newpage
\baselineskip=20pt
\begin{flushleft}
{\Large Figure Captions }
\end{flushleft}

\noindent
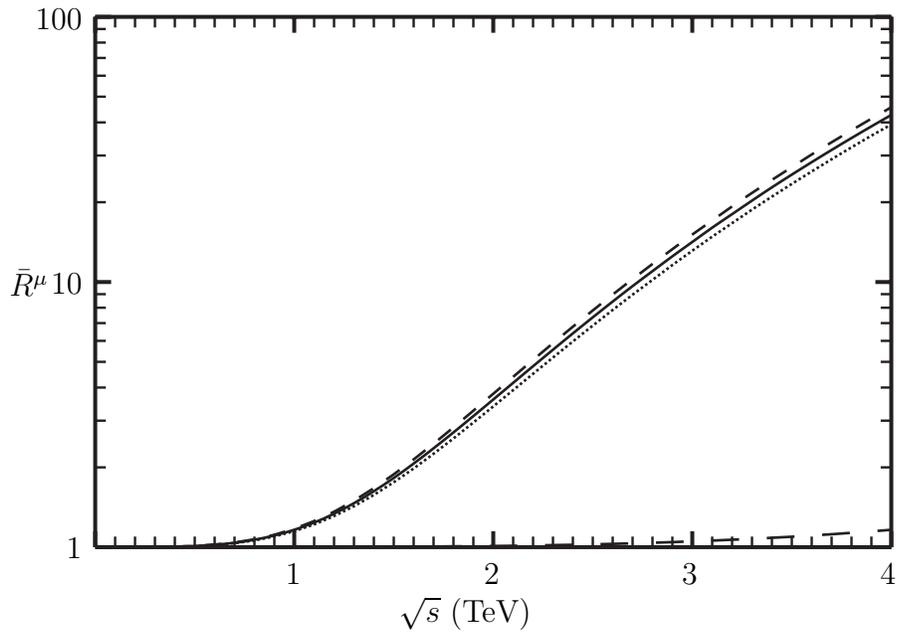
Fig. 2. The NC $\mu^{\pm}$ cross section normalized to its QED 
counterpart is shown as a function of $\sqrt{s}$ at the TESLA site. Other 
parameters are, $\rho=0$ (dotted), $\pi/4$ (solid) and $\pi/2$ (dashed) 
with $\Lambda_{\rm NC}=1$ TeV, and $\rho=\pi/4$ with 
$\Lambda_{\rm NC}=4$ TeV (long-dashed).

\noindent
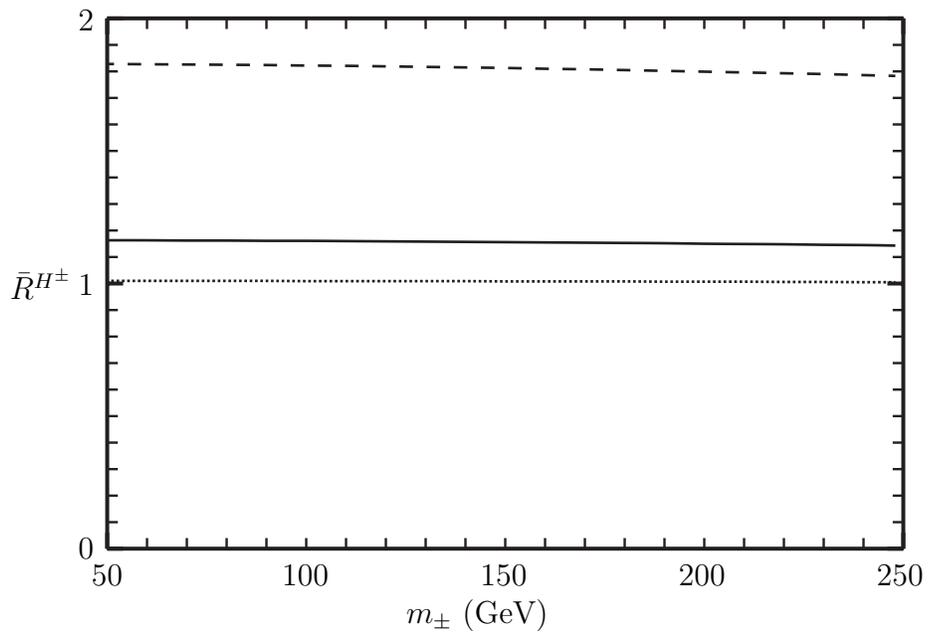
Fig. 3. The NC $H^{\pm}$ cross section normalized to its QED 
counterpart is shown as a function of its mass at the TESLA site and 
for $\sqrt{s}=0.5$ (dotted), $1.0$ (solid) or $1.5$ (dashed) TeV 
and $\Lambda_{\rm NC}=1$ TeV, $\rho=\pi/4$. 

\noindent
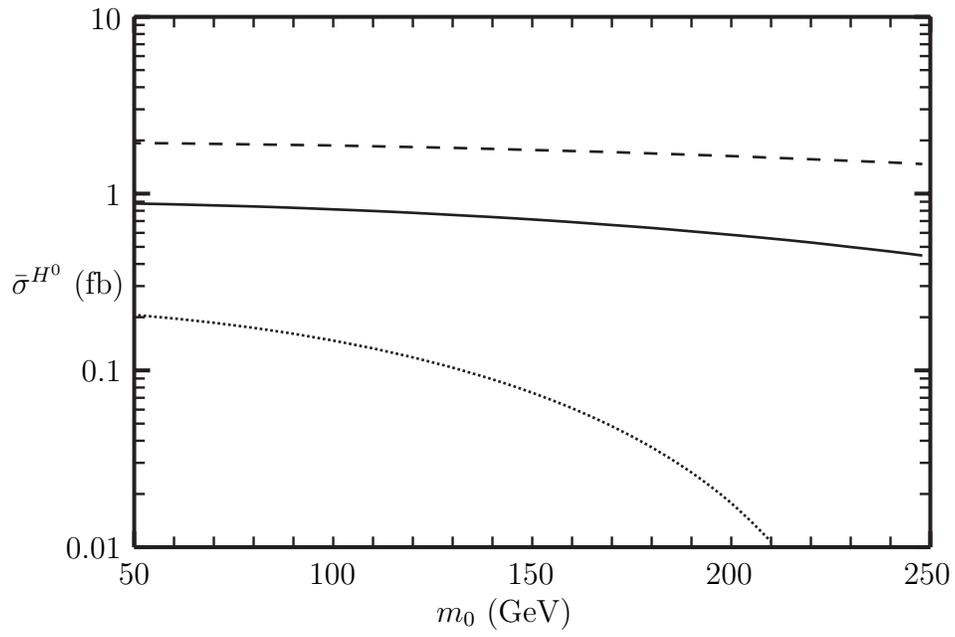
Fig. 4. The NC $H^0$ cross section is shown as a function of its mass 
using the same parameters as in Fig. $3$. 

\noindent
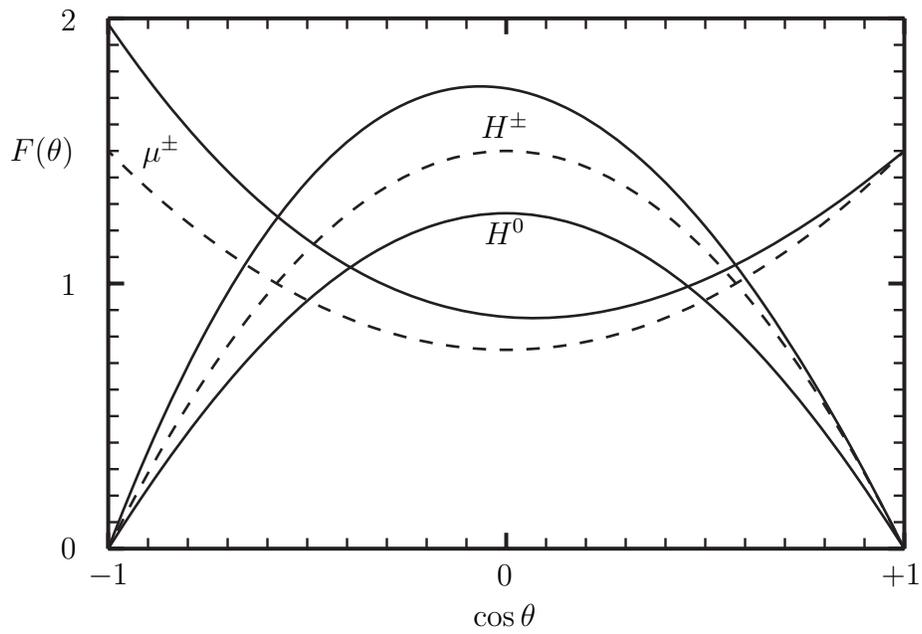
Fig. 5. The normalized and averaged $\theta$ distributions are shown 
for the NC processes (solid) and compared to their QED counterparts
for $\mu^{\pm}$ and $H^{\pm}$ production (dashed). 
$\rho=\pi/4$, $\Lambda_{\rm NC}=\sqrt{s}=1$ TeV, 
$m_{\pm}=m_0=150$ GeV. 

\noindent
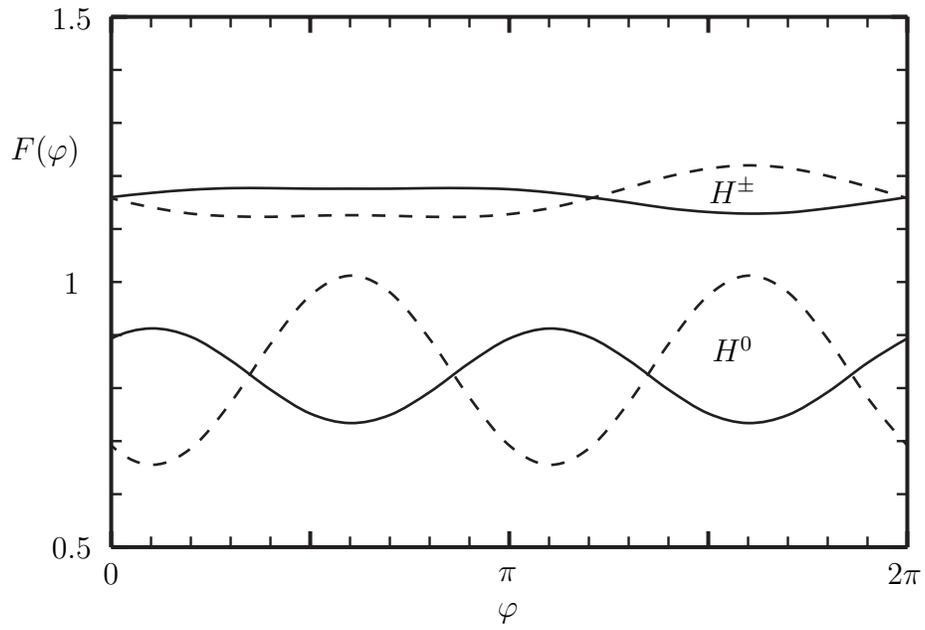
Fig. 6. The normalized and averaged $\varphi$ distributions are shown 
for $H^{\pm}$ and $H^0$ production at $\rho=\pi/4$ (dashed) or 
$\pi/3$ (solid). The case for $\mu^{\pm}$ production is close to 
$H^{\pm}$ and thus not shown. Other parameters are the same as in 
Fig. $5$. 

\noindent
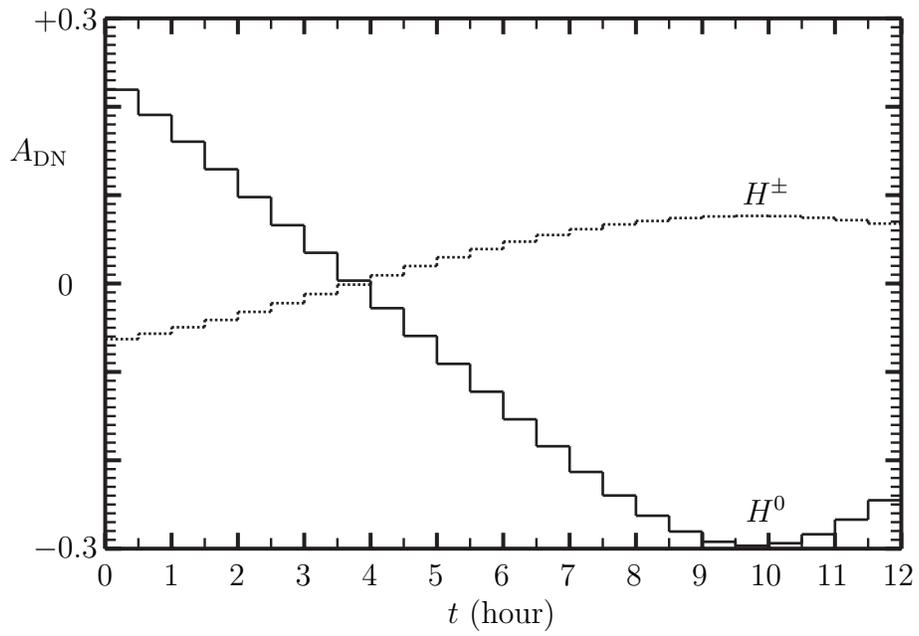
Fig. 7. The histograms of the day-night asymmetry $A_{\rm DN}$ are 
shown as a function of time $t$ using the same parameters as in 
Fig. $5$. 

\newpage
\begin{center}
\begin{picture}(350,300)(0,0)

\SetOffset(40,50)\SetWidth{1.}
\LinAxis(0,0)(300,0)(4,10,5,0,1.5)
\LinAxis(0,200)(300,200)(4,10,-5,0,1.5)
\LogAxis(0,0)(0,200)(2,-5,0,1.5)
\LogAxis(300,0)(300,200)(2,5,0,1.5)
\Text(75,-10)[]{$1$}\Text(150,-10)[]{$2$}\Text(225,-10)[]{$3$}
\Text(300,-10)[]{$4$}\Text(140,-25)[]{$\sqrt{s}$ (TeV)}
\Text(-5,0)[r]{$1$}\Text(-5,100)[r]{$10$}
\Text(-5,200)[r]{$100$}\Text(-25,100)[]{$\bar{R}^{\mu}$}
\Text(140,-50)[]{Figure $2$}
\DashCurve{(7.500,0.000)(15.00,0.000)(22.50,0.043)(30.00,0.173)
(37.50,0.389)(45.00,0.817)(52.50,1.535)(60.00,2.571)
(67.50,4.060)(75.00,6.069)(82.50,8.600)(90.00,11.76)
(97.50,15.47)(105.0,19.72)(112.5,24.50)(120.0,29.68)
(127.5,35.23)(135.0,41.02)(142.5,47.01)(150.0,53.09)
(157.5,59.23)(165.0,65.39)(172.5,71.52)(180.0,77.59)
(187.5,83.56)(195.0,89.45)(202.5,95.22)(210.0,100.8)
(217.5,106.4)(225.0,111.8)(232.5,117.1)(240.0,122.2)
(247.5,127.3)(255.0,132.2)(262.5,137.0)(270.0,141.7)
(277.5,146.3)(285.0,150.8)(292.5,155.2)(300.0,159.4)}{1}
\Curve{(7.500,0.000)(15.00,0.000)(22.50,0.043)(30.00,0.173)
(37.50,0.432)(45.00,0.902)(52.50,1.661)(60.00,2.816)
(67.50,4.414)(75.00,6.557)(82.50,9.272)(90.00,12.64)
(97.50,16.58)(105.0,21.11)(112.5,26.10)(120.0,31.53)
(127.5,37.29)(135.0,43.28)(142.5,49.44)(150.0,55.69)
(157.5,61.98)(165.0,68.25)(172.5,74.47)(180.0,80.63)
(187.5,86.68)(195.0,92.63)(202.5,98.46)(210.0,104.1)
(217.5,109.7)(225.0,115.1)(232.5,120.5)(240.0,125.6)
(247.5,130.7)(255.0,135.7)(262.5,140.5)(270.0,145.2)
(277.5,149.8)(285.0,154.3)(292.5,158.7)(300.0,163.0)}
\DashCurve{(7.500,0.000)(15.00,0.000)(22.50,0.043)(30.00,0.173)
(37.50,0.475)(45.00,0.987)(52.50,1.786)(60.00,2.978)
(67.50,4.688)(75.00,6.966)(82.50,9.864)(90.00,13.38)
(97.50,17.55)(105.0,22.27)(112.5,27.48)(120.0,33.10)
(127.5,39.02)(135.0,45.17)(142.5,51.48)(150.0,57.86)
(157.5,64.25)(165.0,70.62)(172.5,76.93)(180.0,83.15)
(187.5,89.27)(195.0,95.26)(202.5,101.1)(210.0,106.8)
(217.5,112.4)(225.0,117.9)(232.5,123.2)(240.0,128.5)
(247.5,133.6)(255.0,138.5)(262.5,143.3)(270.0,148.1)
(277.5,152.7)(285.0,157.2)(292.5,161.6)(300.0,165.9)}{6}
\DashCurve{(7.500,0.000)(15.00,0.000)(22.50,0.000)(30.00,0.000)
(37.50,0.000)(45.00,0.000)(52.50,0.000)(60.00,0.000)
(67.50,0.000)(75.00,0.043)(82.50,0.043)(90.00,0.043)
(97.50,0.086)(105.0,0.086)(112.5,0.130)(120.0,0.173)
(127.5,0.216)(135.0,0.302)(142.5,0.346)(150.0,0.432)
(157.5,0.518)(165.0,0.646)(172.5,0.774)(180.0,0.902)
(187.5,1.072)(195.0,1.241)(202.5,1.452)(210.0,1.661)
(217.5,1.911)(225.0,2.201)(232.5,2.489)(240.0,2.816)
(247.5,3.140)(255.0,3.542)(262.5,3.941)(270.0,4.414)
(277.5,4.883)(285.0,5.422)(292.5,5.956)(300.0,6.557)}{10}
\end{picture}
\end{center}
\begin{center}
\begin{picture}(350,300)(0,0)

\SetOffset(40,50)\SetWidth{1.}
\LinAxis(0,0)(300,0)(1,20,5,0,1.5)
\LinAxis(0,200)(300,200)(1,20,-5,0,1.5)
\LinAxis(0,0)(0,200)(2,10,-5,0,1.5)
\LinAxis(300,0)(300,200)(2,10,5,0,1.5)
\Text(0,-10)[]{$50$}\Text(75,-10)[]{$100$}
\Text(150,-10)[]{$150$}\Text(225,-10)[]{$200$}
\Text(300,-10)[]{$250$}\Text(140,-25)[]{$m_{\pm}$ (GeV)}
\Text(-5,0)[r]{$0$}\Text(-5,100)[r]{$1$}
\Text(-5,200)[r]{$2$}\Text(-25,100)[]{$\bar{R}^{H^{\pm}}$}
\Text(140,-50)[]{Figure $3$}
\DashCurve{(0.000,101.00)(15.00,101.00)(30.00,101.00)(45.00,101.00)
(60.00,101.00)(75.00,100.90)(90.00,100.90)(105.0,100.90)
(120.0,100.90)(135.0,100.90)(150.0,100.80)(165.0,100.80)
(180.0,100.80)(195.0,100.80)(210.0,100.70)(225.0,100.70)
(240.0,100.70)(255.0,100.60)(270.0,100.60)(285.0,100.50)
(297.0,100.50)}{1}
\Curve{(0.000,116.30)(15.00,116.30)(30.00,116.20)(45.00,116.20)
(60.00,116.10)(75.00,116.10)(90.00,116.00)(105.0,115.90)
(120.0,115.80)(135.0,115.70)(150.0,115.60)(165.0,115.50)
(180.0,115.40)(195.0,115.30)(210.0,115.20)(225.0,115.00)
(240.0,114.90)(255.0,114.80)(270.0,114.60)(285.0,114.50)
(297.0,114.30)}
\DashCurve{(0.000,182.80)(15.00,182.70)(30.00,182.60)(45.00,182.50)
(60.00,182.40)(75.00,182.20)(90.00,182.10)(105.0,181.90)
(120.0,181.70)(135.0,181.50)(150.0,181.30)(165.0,181.00)
(180.0,180.80)(195.0,180.50)(210.0,180.20)(225.0,179.90)
(240.0,179.60)(255.0,179.30)(270.0,179.00)(285.0,178.60)
(297.0,178.30)}{5}
\end{picture}
\end{center}
\begin{center}
\begin{picture}(350,300)(0,0)

\SetOffset(40,50)\SetWidth{1.}
\LinAxis(0,0)(300,0)(1,20,5,0,1.5)
\LinAxis(0,200)(300,200)(1,20,-5,0,1.5)
\LogAxis(0,0)(0,200)(3,-5,0,1.5)
\LogAxis(300,0)(300,200)(3,5,0,1.5)
\Text(0,-10)[]{$50$}\Text(75,-10)[]{$100$}
\Text(150,-10)[]{$150$}\Text(225,-10)[]{$200$}
\Text(300,-10)[]{$250$}\Text(140,-25)[]{$m_0$ (GeV)}
\Text(-5,0)[r]{$0.01$}\Text(-5,66.7)[r]{$0.1$}
\Text(-5,133.3)[r]{$1$}\Text(-5,200)[r]{$10$}
\Text(-25,100)[]{$\bar{\sigma}^{H^0}~({\rm fb})$}
\Text(140,-50)[]{Figure $4$}
\DashCurve{(0.000,87.59)(15.00,86.25)(30.00,84.63)(45.00,82.71)
(60.00,80.49)(75.00,77.91)(90.00,74.96)(105.0,71.58)
(120.0,67.71)(135.0,63.30)(150.0,58.24)(165.0,52.40)
(180.0,45.62)(195.0,37.66)(210.0,28.17)(225.0,16.59)
(240.0,2.039)}{1}
\Curve{(0.000,129.6)(15.00,129.3)(30.00,128.9)(45.00,128.5)
(60.00,128.0)(75.00,127.4)(90.00,126.8)(105.0,126.1)
(120.0,125.3)(135.0,124.5)(150.0,123.6)(165.0,122.6)
(180.0,121.5)(195.0,120.4)(210.0,119.1)(225.0,117.8)
(240.0,116.4)(255.0,114.9)(270.0,113.2)(285.0,111.5)
(297.0,110.0)}
\DashCurve{(0.000,152.4)(15.00,152.3)(30.00,152.1)(45.00,151.9)
(60.00,151.7)(75.00,151.5)(90.00,151.2)(105.0,150.9)
(120.0,150.6)(135.0,150.2)(150.0,149.8)(165.0,149.4)
(180.0,149.0)(195.0,148.5)(210.0,148.0)(225.0,147.5)
(240.0,146.9)(255.0,146.3)(270.0,145.7)(285.0,145.1)
(297.0,144.5)}{5}
\end{picture}
\end{center}
\begin{center}
\begin{picture}(350,300)(0,0)

\SetOffset(40,50)\SetWidth{1.}
\LinAxis(0,0)(300,0)(2,10,5,0,1.5)
\LinAxis(0,200)(300,200)(2,10,-5,0,1.5)
\LinAxis(0,200)(0,0)(2,10,5,0,1.5)
\LinAxis(300,0)(300,200)(2,10,5,0,1.5)
\Text(0,-10)[]{$-1$}\Text(150,-10)[]{$0$}
\Text(300,-10)[]{$+1$}\Text(150,-25)[]{$\cos\theta$}
\Text(-15,0)[]{$0$}\Text(-15,100)[]{$1$}
\Text(-15,200)[]{$2$}\Text(-25,150)[]{$F(\theta)$}
\Text(20,150)[]{$\mu^{\pm}$}\Text(150,160)[]{$H^{\pm}$}
\Text(150,120)[]{$H^0$}
\Text(150,-50)[]{Figure $5$}
\Curve{(0.000,197.9)(2.280,194.6)(9.044,185.1)(20.10,170.5)
(35.10,152.8)(53.58,134.1)(75.00,116.6)(98.70,102.1)
(123.9,92.13)(150.0,87.37)(176.0,87.85)(201.3,92.96)
(225.0,101.6)(246.4,112.4)(264.9,123.7)(279.9,134.2)
(290.9,142.7)(297.7,148.1)(300.0,150.0)}
\DashCurve{(0.000,150.0)(2.280,147.7)(9.044,141.2)(20.10,131.2)
(35.10,119.0)(53.58,106.0)(75.00,93.75)(98.70,83.77)
(123.9,77.26)(150.0,75.00)(176.0,77.26)(201.3,83.77)
(225.0,93.75)(246.4,106.0)(264.9,119.0)(279.9,131.3)
(290.9,141.2)(297.7,147.7)(300.0,150.0)}{5}
\Curve{(0.000,0.   )(2.280,5.892)(9.044,22.74)(20.10,48.21)
(35.10,78.78)(53.58,110.3)(75.00,138.6)(98.70,160.1)
(123.9,172.2)(150.0,173.6)(176.0,164.5)(201.3,146.3)
(225.0,121.5)(246.4,93.07)(264.9,64.32)(279.9,38.32)
(290.9,17.72)(297.7,4.535)(300.0,0.   )}
\DashCurve{(0.000,0.   )(2.280,4.523)(9.044,17.55)(20.10,37.50)
(35.10,61.98)(53.58,88.02)(75.00,112.5)(98.70,132.5)
(123.9,145.5)(150.0,150.0)(176.0,145.5)(201.3,132.5)
(225.0,112.5)(246.4,88.02)(264.9,61.98)(279.9,37.50)
(290.9,17.55)(297.7,4.523)(300.0,0.   )}{5}
\Curve{(0.000,0.   )(2.280,3.584)(9.044,13.99)(20.10,30.17)
(35.10,50.41)(53.58,72.39)(75.00,93.47)(98.70,110.9)
(123.9,122.4)(150.0,126.5)(176.0,122.4)(201.3,110.9)
(225.0,93.48)(246.4,72.42)(264.9,50.42)(279.9,30.18)
(290.9,13.99)(297.7,3.584)(300.0,0.   )}
\end{picture}
\end{center}
\begin{center}
\begin{picture}(350,300)(0,0)

\SetOffset(40,50)\SetWidth{1.}
\LinAxis(0,0)(300,0)(4,5,5,0,1.5)
\LinAxis(0,200)(300,200)(4,5,-5,0,1.5)
\LinAxis(0,200)(0,0)(1,10,5,0,1.5)
\LinAxis(300,0)(300,200)(1,10,5,0,1.5)
\Text(0,-10)[]{$0$}\Text(150,-10)[]{$\pi$}
\Text(300,-10)[]{$2\pi$}\Text(150,-25)[]{$\varphi$}
\Text(-15,0)[]{$0.5$}\Text(-15,100)[]{$1$}
\Text(-15,200)[]{$1.5$}\Text(-25,150)[]{$F(\varphi)$}
\Text(235,135)[]{$H^{\pm}$}\Text(235,75)[]{$H^0$}
\Text(140,-50)[]{Figure $6$}
\DashCurve{(0.000,131.6)(15.00,128.2)(30.00,125.8)(45.00,124.8)
(60.00,124.6)(75.00,125.0)(90.00,125.2)(105.0,125.0)
(120.0,124.6)(135.0,124.6)(150.0,125.6)(165.0,127.8)
(180.0,131.4)(195.0,135.6)(210.0,139.8)(225.0,142.8)
(240.0,144.0)(255.0,143.0)(270.0,140.0)(285.0,136.0)
(300.0,131.6)}{5}
\DashCurve{(0.000,38.38)(15.00,31.08)(30.00,37.24)(45.00,54.50)
(60.00,76.56)(75.00,94.88)(90.00,102.4)(105.0,96.12)
(120.0,78.52)(135.0,56.43)(150.0,38.38)(165.0,31.06)
(180.0,37.22)(195.0,54.52)(210.0,76.56)(225.0,94.92)
(240.0,102.4)(255.0,96.14)(270.0,78.54)(285.0,56.43)
(300.0,38.38)}{5}
\Curve{(0.000,132.0)(15.00,133.6)(30.00,134.8)(45.00,135.4)
(60.00,135.4)(75.00,135.2)(90.00,135.2)(105.0,135.2)
(120.0,135.4)(135.0,135.4)(150.0,135.0)(165.0,133.8)
(180.0,132.0)(195.0,130.0)(210.0,127.8)(225.0,126.4)
(240.0,125.8)(255.0,126.2)(270.0,127.8)(285.0,129.8)
(300.0,132.0)}
\Curve{(0.000,78.70)(15.00,82.50)(30.00,79.35)(45.00,70.43)
(60.00,59.42)(75.00,50.39)(90.00,46.80)(105.0,49.84)
(120.0,58.48)(135.0,69.48)(150.0,78.68)(165.0,82.46)
(180.0,79.32)(195.0,70.48)(210.0,59.44)(225.0,50.42)
(240.0,46.80)(255.0,49.84)(270.0,58.48)(285.0,69.50)
(300.0,78.70)}
\end{picture}
\end{center}
\begin{center}
\begin{picture}(350,300)(0,0)

\SetOffset(40,50)\SetWidth{1.}
\LinAxis(0,0)(300,0)(12,2,5,0,1.5)
\LinAxis(0,200)(300,200)(12,2,-5,0,1.5)
\LinAxis(0,200)(0,0)(6,10,5,0,1.5)
\LinAxis(300,0)(300,200)(6,10,5,0,1.5)
\Text(0,-10)[]{$0$}\Text(25,-10)[]{$1$}\Text(50,-10)[]{$2$}
\Text(75,-10)[]{$3$}\Text(100,-10)[]{$4$}\Text(125,-10)[]{$5$}
\Text(150,-10)[]{$6$}\Text(175,-10)[]{$7$}\Text(200,-10)[]{$8$}
\Text(225,-10)[]{$9$}\Text(250,-10)[]{$10$}\Text(275,-10)[]{$11$}
\Text(300,-10)[]{$12$}\Text(150,-25)[]{$t$ (hour)}
\Text(-15,0)[]{$-0.3$}\Text(-15,100)[]{$0$}\Text(-15,200)[]{$+0.3$}
\Text(-25,150)[]{$A_{\rm DN}$}
\Text(250,135)[]{$H^{\pm}$}\Text(250,15)[]{$H^0$}
\Text(140,-50)[]{Figure $7$}
\DashLine(0.000,79.03)(12.50,79.03){1}\DashLine(12.50,79.03)(12.50,81.09){1}
\DashLine(12.50,81.09)(25.00,81.09){1}\DashLine(25.00,81.09)(25.00,83.51){1}
\DashLine(25.00,83.51)(37.50,83.51){1}\DashLine(37.50,83.51)(37.50,86.26){1}
\DashLine(37.50,86.26)(50.00,86.26){1}\DashLine(50.00,86.26)(50.00,89.31){1}
\DashLine(50.00,89.31)(62.50,89.31){1}\DashLine(62.50,89.31)(62.50,92.59){1}
\DashLine(62.50,92.59)(75.00,92.59){1}\DashLine(75.00,92.59)(75.00,96.04){1}
\DashLine(75.00,96.04)(87.50,96.04){1}\DashLine(87.50,96.04)(87.50,99.59){1}
\DashLine(87.50,99.59)(100.0,99.59){1}\DashLine(100.0,99.59)(100.0,103.1){1}
\DashLine(100.0,103.1)(112.5,103.1){1}\DashLine(112.5,103.1)(112.5,106.6){1}
\DashLine(112.5,106.6)(125.0,106.6){1}\DashLine(125.0,106.6)(125.0,109.9){1}
\DashLine(125.0,109.9)(137.5,109.9){1}\DashLine(137.5,109.9)(137.5,113.0){1}
\DashLine(137.5,113.0)(150.0,113.0){1}\DashLine(150.0,113.0)(150.0,115.8){1}
\DashLine(150.0,115.8)(162.5,115.8){1}\DashLine(162.5,115.8)(162.5,118.3){1}
\DashLine(162.5,118.3)(175.0,118.3){1}\DashLine(175.0,118.3)(175.0,120.5){1}
\DashLine(175.0,120.5)(187.5,120.5){1}\DashLine(187.5,120.5)(187.5,122.3){1}
\DashLine(187.5,122.3)(200.0,122.3){1}\DashLine(200.0,122.3)(200.0,123.6){1}
\DashLine(200.0,123.6)(212.5,123.6){1}\DashLine(212.5,123.6)(212.5,124.7){1}
\DashLine(212.5,124.7)(225.0,124.7){1}\DashLine(225.0,124.7)(225.0,125.3){1}
\DashLine(225.0,125.3)(237.5,125.3){1}\DashLine(237.5,125.3)(237.5,125.5){1}
\DashLine(237.5,125.5)(250.0,125.5){1}\DashLine(250.0,125.5)(250.0,125.4){1}
\DashLine(250.0,125.4)(262.5,125.4){1}\DashLine(262.5,125.4)(262.5,124.8){1}
\DashLine(262.5,124.8)(275.0,124.8){1}\DashLine(275.0,124.8)(275.0,123.9){1}
\DashLine(275.0,123.9)(287.5,123.9){1}\DashLine(287.5,123.9)(287.5,122.6){1}
\DashLine(287.5,122.6)(300.0,122.6){1}
\Line(0.000,173.1)(12.50,173.1)\Line(12.50,173.1)(12.50,163.6)
\Line(12.50,163.6)(25.00,163.6)\Line(25.00,163.6)(25.00,153.5)
\Line(25.00,153.5)(37.50,153.5)\Line(37.50,153.5)(37.50,143.1)
\Line(37.50,143.1)(50.00,143.1)\Line(50.00,143.1)(50.00,132.6)
\Line(50.00,132.6)(62.50,132.6)\Line(62.50,132.6)(62.50,122.0)
\Line(62.50,122.0)(75.00,122.0)\Line(75.00,122.0)(75.00,111.6)
\Line(75.00,111.6)(87.50,111.6)\Line(87.50,111.6)(87.50,101.1)
\Line(87.50,101.1)(100.0,101.1)\Line(100.0,101.1)(100.0,90.70)
\Line(100.0,90.70)(112.5,90.70)\Line(112.5,90.70)(112.5,80.23)
\Line(112.5,80.23)(125.0,80.23)\Line(125.0,80.23)(125.0,69.70)
\Line(125.0,69.70)(137.5,69.70)\Line(137.5,69.70)(137.5,59.20)
\Line(137.5,59.20)(150.0,59.20)\Line(150.0,59.20)(150.0,48.80)
\Line(150.0,48.80)(162.5,48.80)\Line(162.5,48.80)(162.5,38.63)
\Line(162.5,38.63)(175.0,38.63)\Line(175.0,38.63)(175.0,28.93)
\Line(175.0,28.93)(187.5,28.93)\Line(187.5,28.93)(187.5,20.06)
\Line(187.5,20.06)(200.0,20.06)\Line(200.0,20.06)(200.0,12.43)
\Line(200.0,12.43)(212.5,12.43)\Line(212.5,12.43)(212.5,6.466)
\Line(212.5,6.466)(225.0,6.466)\Line(225.0,6.466)(225.0,2.600)
\Line(225.0,2.600)(237.5,2.600)\Line(237.5,2.600)(237.5,1.100)
\Line(237.5,1.100)(250.0,1.100)\Line(250.0,1.100)(250.0,2.066)
\Line(250.0,2.066)(262.5,2.066)\Line(262.5,2.066)(262.5,5.433)
\Line(262.5,5.433)(275.0,5.433)\Line(275.0,5.433)(275.0,10.96)
\Line(275.0,10.96)(287.5,10.96)\Line(287.5,10.96)(287.5,18.26)
\Line(287.5,18.26)(300.0,18.26)
\end{picture}
\end{center}


\newpage
\baselineskip=20pt
\begin{thebibliography}{30}

\bibitem{string}
A. Connes, M. R. Douglas, A. Schwarz, 
{\it Noncommutative geometry and matrix theory: compactification on tori}, 
\JHEP 02(1998)003 [hep-th/9711162]; 
M. R. Douglas, C. Hull, 
{\it D-branes and the noncommutative torus}, 
{\it ibid}. 02(1998)008 [hep-th/9711165];
C.-S. Chu, P.-M. Ho, 
{\it Noncommutative open string and D-brane}, 
\NP B550(1999)151 [hep-th/9812219]; 
{\it Constrained quantization of open string in background $B$ field and 
noncommutative D-brane}, 
{\it ibid}. B568(2000)447 [hep-th/9906192]; 
V. Schomerus, 
{\it D-branes and deformation quantization}, \JHEP 06 (1999) 030 
[hep-th/9903205]; 
N. Seiberg, E. Witten, 
{\it String theory and noncommutative geometry}, 
{\it ibid}. 09(1999)032 [hep-th/9908142].

\bibitem{tev}N. Arkani-Hamed, S. Dimopoulos, G. Dvali, 
{\it The hierarchy problem and new dimensions at a millimeter}, 
\PL B429(1998)263 [hep-ph/9803315]; 
{\it Phenomenology, astrophysics and cosmology of theories with submillimeter 
dimensions and TeV scale quantum gravity}, 
\PR D59(1999)086004 [hep-ph/9807344]; 
I. Antoniadis, N. Arkani-Hamed, S. Dimopoulos, G. Dvali, 
{\it New dimensions at a millimeter to a fermi and superstrings at a TeV}, 
\PL B436(1998)257 [hep-ph/9804398].

\bibitem{hewett}J. L. Hewett, F. J. Petriello, T. G. Rizzo, 
{\it Signals for noncommutative interactions at linear colliders}, 
\PR D64(2001)075012 [hep-ph/0010354].

\bibitem{liao01}H. Grosse, Y. Liao, 
{\it Pair production of neutral Higgs bosons through noncommutative 
QED interactions at linear colliders}, 
\PR D64(2001)115007 [hep-ph/0105090].

\bibitem{kamoshita}J. Kamoshita, 
{\it Probing noncommutative space-time in the laboratory frame}, 
hep-ph/0206223.

\bibitem{high}P. Mathews, 
{\it Compton scattering in noncommutative spacetime at the NLC}, 
\PR D63(2001)075007 [hep-ph/0011332]; 
S. Baek, D. K. Ghosh, X.-G. He, W.-Y. P. Hwang, 
{\it Signatures of noncommutative QED at photon colliders}, 
{\it ibid}. D64(2001)056001 [hep-ph/0103068]; 
S. Godfrey, M.A. Doncheski, 
{\it Signals for noncommutative QED in $e\gamma$ and $\gamma\gamma$ 
collisions}, 
{\it ibid}. D65(2002)015005 [hep-ph/0108268];
C.E. Carlson, C.D. Carone, 
{\it Discerning noncommutative extra dimensions}, 
{\it ibid}. D65(2002)075007 [hep-ph/0112143]; 
T.M. Aliev, O. Ozcan, M. Savci, 
{\it The $\gamma\gamma\to H^0H^0$ decay in noncommutative quantum 
electrodynamics}, 
hep-ph/0209205.

\bibitem{low}I. Mocioiu, M. Pospelov, R. Roiban, 
{\it Low-energy limits on the antisymmetric tensor field background 
on the brane and on the noncommutativity scale}, 
\PL B489(2000)390 [hep-ph/0005191]; 
M. Chaichian, M. M. Sheikh-Jabbari, A. Tureanu,
{\it Hydrogen atom spectrum and the Lamb shift in nocommutative QED}, 
\PRL 86, 2716 (2001) [hep-th/0010175]; 
H. Grosse, Y. Liao, 
{\it Anomalous C violating three photon decay of the neutral pion in 
noncommutative quantum electrodynamics},
\PL B520(2001)63 [hep-ph/0104260]; 
S.M. Carroll, {\it et al}, 
{\it Noncommutative field theory and Lorentz violation}, 
\PRL 87(2001)141601 [hep-th/0105082]; 
I. Mocioiu, M. Pospelov, R. Roiban, 
{\it Breaking CPT by mixed noncommutativity}, 
\PR D65(2002)107702 [hep-ph/0108136]; 
C. E. Carlson, C.D. Carone, R.F. Lebed, 
{\it Supersymmetric noncommutative QED and Lorentz violation}, 
hep-ph/0209077. 

\bibitem{filk}T. Filk, {\it Divergences in a field theory on quantum space}, 
\PL B376(1996)53. 

\bibitem{others}M. Chaichian, A. Demichev, P. Presnajder, 
{\it Quantum field theory on noncommutative space-times and the persistence 
of ultraviolet divergences}, 
\NP B567(2000)360 [hep-th/9812180]; 
C.P. Martin, D. Sanchez-Ruiz, 
{\it The one loop UV divergent structure of $U(1)$ Yang-Mills theory on 
noncommutative $R^4$}, 
\PRL 83(1999)476 [hep-th/9903077]; 
S. Minwalla, M. V. Raamsdonk and N. Seiberg, 
{\it Noncommutative perturbative dynamics}, 
\JHEP 02(2000)020 [hep-th/9912072];
M. Hayakawa, 
{\it Perturbative analysis on infrared aspects of 
noncommutative QED on $R^4$}, 
\PL B478(2000)394 [hep-th/9912094]; 
{\it Perturbative analysis on infrared and ultraviolet aspects of 
noncommutative QED on $R^4$}, 
hep-th/9912167.

\bibitem{unitarity}
J. Gomis, T. Mehen, 
{\it Space-time noncommutative field theories and unitarity}, 
\NP B591(2000)265 [hep-th/0005129]. 

\bibitem{bahns}D. Bahns, S. Doplicher, K. Fredenhagen, G. Piacitelli, 
{\it On the unitarity problem in space-time noncommutative theories}, 
\PL B533(2002)178 [hep-th/0201222]. 

\bibitem{rim}C. Rim, J. H. Yee, 
{\it Unitarity in space-time noncommutative field theories}, 
hep-th/0205193.

\bibitem{doplicher}S. Doplicher, K. Fredenhagen, J. E. Roberts, 
{\it The quantum structure of spacetime at the Planck scale and quantum 
fields}, 
Commun. Math. Phys. 172(1995)187.

\bibitem{liao}Y. Liao, K. Sibold, 
{\it Time-ordered perturbation theory on noncommutative spacetime: 
basic rules}, 
\EPJ C25(2002)469 [hep-th/0205269]; 
{\it Time-ordered perturbation theory on noncommutative spacetime II: 
unitarity}, 
\EPJ C25(2002)479 [hep-th/0206011]. 

\bibitem{bozkaya}H. Bozkaya, {\it et al}, 
{\it Space-time noncommutative field theories and causality}, 
hep-th/0209253.

\bibitem{nogo}For a recent comprehensive discussion, see for example: 
M. Chaichian, P. Presnajder, M. M. Sheikh-Jabbari, A. Tureanu, 
{\it Noncommutative gauge field theories: a no-go theorem}, 
\PL B526 (2002) 132 [hep-th/0107037]; 
{\it Noncommutative standard model: model building}, 
hep-th/0107055. 
For discussions on the unitarity problem in the latter work, see: 
J. L. Hewett, F. J. Petriello, T. G. Rizzo, 
{\it Noncommutativity and unitarity violation in gauge boson scattering}, 
\PR D66(2002)036001 [hep-ph/0112003]. 

\bibitem{zerwas}P. M. Zerwas, private communication. 

\bibitem{eehh}K. J. F. Gaemers, F. Hoogeveen, 
{\it Higgs boson pair production in $e^+e^-$ reactions}, 
Z. Phys. C26(1984)249; 
A. Djouadi, V. Driesen, C. Junger, 
{\it Loop induced Higgs boson pair production at $e^+e^-$ colliders}, 
\PR D 54(1996)759 [hep-ph/9602341].

\bibitem{froissart}M. Froissart, 
{\it Asymptotic behaviour and subtractions in the Mandelstam 
representation}, 
\PR 123(1961)1053. 

\bibitem{mandelstam}S. Mandelstam, 
{\it Determination of the pion-nucleon scattering amplitude from 
dispersion relations and unitarity. General theory}, 
\PR 112(1958)1344.

\bibitem{martin}A. Martin, 
{\it Unitarity and high-energy behavior of scattering amplitudes}, 
\PR 129(1963)1432. 

\bibitem{dispersion}Y. Liao, K. Sibold, 
{\it Spectral representation and dispersion relations in field theory 
on noncommutative space}, 
\PL B549(2002)352 [hep-th/0209221].

\bibitem{newton}See for example: R.G. Newton, {\it Scattering theory of 
waves and particles} (Springer-Verlag, 1982).

\end{thebibliography}
\end{document}